
\input harvmac

\input epsf
\ifx\epsfbox\UnDeFiNeD\message{(NO epsf.tex, FIGURES WILL BE IGNORED)}
\def\figin#1{\vskip2in}
\else\message{(FIGURES WILL BE INCLUDED)}\def\figin#1{#1}\fi
\def\ifig#1#2#3{\xdef#1{fig.~\the\figno}
\goodbreak\midinsert\figin{\centerline{#3}}%
\smallskip\centerline{\vbox{\baselineskip12pt
\advance\hsize by -1truein\noindent{\bf Fig.~\the\figno:} #2}}
\bigskip\endinsert\global\advance\figno by1}

\Title{\vbox{\baselineskip12pt\hbox{hep-th/9302070}\hbox{CALT-68-1852}}}
{\vbox{\centerline{Nonsingular 2-D Black Holes}
\smallskip
\centerline{and Classical String Backgrounds}}}
\bigskip
\centerline{Piljin Yi\footnote{*}{e-mail address:
piljin@theory3.caltech.edu}\footnote{\dag}{Supported in part
by U.S. Dept. of Energy under Contract no. DEAC-03-81ER40050}}
\smallskip
\centerline{\it California Institute of Technology, Pasadena, CA 91125}
\noindent
\bigskip
\baselineskip 18pt
\noindent
We study a string-inspired classical 2-D effective field theory
with {\it nonsingular} black holes as well as Witten's black hole among
its static solutions. By a dimensional reduction, the static  solutions
are related to the  $(SL(2,R)_{k}\otimes U(1))/U(1)$ coset model, or more
precisely its $O\bigl((\alpha')^{0}\bigr)$ approximation known as the
3-D charged black  string. The 2-D effective action possesses a
propagating degree of freedom, and the dynamics are highly nontrivial. A
collapsing shell is shown to bounce into another universe without creating
a curvature singularity on its path, and the potential instability of the
Cauchy horizon is found to be irrelevent in that some of the infalling
observers never approach the Cauchy horizon.
Finally a $SL(2,R)_{k}/U(1)$ nonperturbative coset metric, found and
advocated by R. Dijkgraaf et.al., is shown to be nonsingular and
to coincide with one of the charged spacetimes found above.
Implications of all these geometries
are discussed in connection with  black hole evaporation.

\Date{February 1993}

\newsec{Introduction}
\lref\HAWK{S.W. Hawking, Comm.Math.Phys.43 (1975) 199.}
\lref\HAWKINF{S.W. Hawking, Phys.Rev.D14 (1976) 2460.}
\lref\WITT{E. Witten, Phys.Rev.D44 (1991) 314.}
\lref\CGHS{C.G. Callan, S.B. Giddings, J.A. Harvey and A. Strominger
           Phys.Rev.D45 (1992) R1005.}
\lref\BSTRING{J.H. Horne and G.T. Horowitz,
              Phys.Rev.Lett. 68 (1992) 568; Nuc.Phys.B368 (1992) 444.}
\lref\DUAL{T. Buscher, Phys.Lett.B201 (1988) 466; B194 (1987) 59.}
\lref\POISSON{E. Poisson and W. Israel, Phys.Rev.D41 (1990) 1796.}
\lref\ORI{A. Ori, Phys.Rev.Lett. 67 (1991) 1796.}
\lref\ISRAEL{V. De La Cruz and W. Israel, Il Nuovo Cimento LI A N.3
             (1967) 745.}
\lref\PRICE{R.H Price, Phys.Rev.D5 (1972) 2419.}
\lref\SUSS{J.G. Russo, L. Susskind and L.Thorlacius, Phys.Rev.D46 (1992)
           3444; Phys.Lett.B292 (1992) 13.}
\lref\BILAL{A. Bilal and C. Callan, {\it{Liouville models of
            black hole evaporation}}, PUPT-1320-Rev.}
\lref\GHOST{A. Strominger, {\it{Fadeev-Popov ghosts and 1+1 dimensional
            black hole evaporation}}, UCSBTH-92-18.}
\lref\REMNANT{T. Banks, M. O'Loughlin and A. Strominger, {\it{Black hole
             remnants and the information puzzle}}, RU-92-40; T. Banks
             and M. O'Loughlin, {\it{Classical and quantum production
             of cornucopions at energy below $10^{18}$ Gev}}, RU-92-14;
             T. Banks, A. Dabolkar, M.R. Douglas and M. O'Loughlin,
             Phys.Rev.D45 (1992) 3607.}
\lref\GIDD{S.B. Giddings and A. Strominger, Phys.Rev.D46 (1992) 627.}
\lref\PRESK{J. Preskill, {\it{Do black holes destroy information?}},
            CALT-68-1819.}
\lref\TRIVEDI{S.P. Trivedi, {\it{Semiclassical extremal black holes}},
           CALT-68-1833.}
\lref\NEW{T. Banks and M. O'Loughlin, {\it{Nonsingular Lagrangian
          for two-dimensional black holes}}, RU-92-61.}
\lref\VVD{R. Dijkgraaf, E. Verlinde and H. Verlinde, Nuc.Phys.B371
          (1992) 269.}
\lref\TSEY{A.A. Tseytlin, {\it{Effective action of guaged WZW model
          and exact string solutions}}, Imperial/TP/92-93/10.}
\lref\SIGM{C.G. Callan, D. Friedan, E.J. Martinec and M.J. Perry,
           Nuc.Phys.B262 (1985) 593.}

The longstanding problem of black hole evaporation has been recently
revitalized in the context of 2-D dilaton gravity \CGHS, either on its own
or as an effective S-wave sector theory of the 4-D black holes \GIDD. In 2-D,
the coupling to conformal matter, hence the Hawking radiation,
can be conveniently represented by the Polyakov-Liouville action of
appropriate coefficient, enabling a systematic study of the gravitational
backreaction \CGHS\SUSS. But most of these studies are again plagued by
curvature singularities, hidden or naked, and the inability to handle
the singularities properly clouds any conclusion concerning the fate of
the black holes \SUSS\BILAL\GHOST.

The question arises, then, whether the singularity is an essential feature
of black hole physics. For example the Hawking radiation is, strictly
speaking, a property of event horizons, and singularities appear only as a
result of the gravitational field equation. Is there a physically sensible
theory, solutions of which are nonsingular spacetimes but with horizons?
The answer is yes. In this paper, we present such an 2-D effective field
theory, static solutions of which are related to certain WZW coset models.

While WZW coset models provide us with string theories with interesting
target spacetimes, it is often very difficult to study the dynamics since
we have to solve the full string theory around nontrivial backgrounds.
An easy way out is to abandon the exact model for an effective
(approximate) local field theory by using the $\alpha'$ expansion of the sigma
model \SIGM. Higher order corrections can be controlled by introducing
extended supersymmetry in some cases, but not always. In particular,
the $\alpha'$ expansion is also an expansion in the curvature of the
target metric and a curvature singularity can be a signal of  big
higher order and nonperturbative corrections.

Apparently this is exactly what happens with Witten's black hole
as an $O\bigl((\alpha')^{0}\bigr)$ approximation to the $SL(2,R)_{k}/U(1)$
coset model. So far, two independent attempts to find a nonperturbative
{\it exact} geometry have yielded an identical static metric \VVD\TSEY, and,
quite surprisingly, this new metric has the causal structure of a
{\it nonsingular}  multi-universe spacetime with horizons, as will be
shown in the final section  of this paper.

Furthermore we have found a string inspired local field
theory in 2-D, whose charged static solutions are mostly  of the same causal
structure as this {\it exact} coset metric.
The action is obtained by a dimensional reduction of the 3-D
string effective action of the gravity multiplet and the static solutions
are dimensionally reduced versions of the 3-D black strings of Horne et.al.
\BSTRING. (These 3-D black strings are
$O\bigl((\alpha')^{0}\bigr)$ approximations to the $(SL(2,R)_{k}\otimes U(1))/
U(1)$ coset model.)

In short, we have not only an exact classical string background
whose metric is a nonsingular black hole but also a local field theory,
with essentially same type of static solutions, which can serve as a
dynamical toy model.  The form of the field theoretical solutions
is identical to that of the $SL(2,R)_{k}/U(1)$ {\it exact} coset background up
to an additive shift of one parameter. Not only the causal structures
but the  behaviour of the dilatons are the same. This presents us with
a unique opportunity as well as a motivation to
study these new kind of black holes.
In this paper we study the classical properties of the 2-D effective
field theory for the most part and detailed discussions on the {\it exact}
background is postponed until at the end.

In section 2, we introduce
the 2-D effective field theory and the much advertised nonsingular
2-D spacetimes. We start with a compactified charged black string.
The geodesic equation of motion is studied for radial motion,
and the Penrose diagrams of the 2-D part of the metric are shown.
After the calculation of the Hawking temperature, we elaborate on the
relationship of these new spacetimes to Witten's black hole, with
emphasis on the duality transformation used for the construction of the
black string from Witten's black hole.

We devote the next two sections to the dynamics of the 2-D effective
theory, in particular, the stability of the static solutions under the
process of a gravitational collapse. In section 3, we investigate the
reponse of the geometry to a collapsing shell of massless matter. Even
though we are unable to solve the full partial differential equations
governing the process, the massless nature of the gravitating source
allows us to obtain useful information such as the difference of the
curvature across the shell. Crucial to the calculation here is the
asymptotic behaviour of the gravitational perturbation. We added an
appendix at the end of the paper to derive the neccessary
information. In section 4, we examine the instability
of the Cauchy horizon and the consequences.

  Section 5 is devoted to some of the  unresolved issues of the
gravitational collapse as well as implications of the results we
obtained in the earlier sections. In particular, the {\it exact}
metric of the $SL(2,R)_{k}/U(1)$, found in \VVD\ and recently
rediscovered in an entirely different approach by A.A. Tseytlin \TSEY,
is shown to have the same geometry as one of the nonextremal
(hence nonsingular) solutions. Also discussed are nonperturbative
black string solutions. Based on this new set of charged spacetimes,
we speculate on the real nature of the 2-D black holes.

\newsec{Compactified Black Strings as Geodesically Complete 2-D Spacetimes}

 We start with the charged black string solution by Horne et.al.\BSTRING,

\eqn\actth{S^{(3)}=\int \sqrt{-g^{(3)}} e^{-2\phi} (R^{(3)}+ 4 (\nabla\phi)^2
                  +4 \lambda^2-{1 \over 12}H^2)}
\eqn\metth{g^{(3)}=-(1-{m \over r})\,dt^2 +
                  {1 \over 4\lambda^2 (r-m) (r-q)}\,dr^2+
                  (1-{q \over r}) a^2 \,d\theta^2}
$${e^{-2\phi}={r \over \lambda}}$$

\noindent
and consider the case $\theta$ periodic with period $2\pi$. Then $a$
is the asymptotic radius of the internal circle. We define $q\equiv Q^2/m$
where $Q$ is the axionic
charge associated with the antisymmetric 2-tensor.
The solution \metth\ is actually $O\bigl((\alpha')^{0}\bigl)$ approximation
to the coset model $(SL(2,R)_{k}\otimes U(1))/U(1)$ and $q=0$ limit
corresponds to $(SL(2,R)_{k}/U(1))\otimes S^{1}$, the Witten's coset
model multiplied by a circle of fixed radius.
For $m >q > 0$, the metric describes singularities at $r=0$
hidden inside two distinct sets of horizons at $r = m,q$. The leading
divergence of the curvature near $r=0$ is $\sim -(1/r^2)$.
The causal structure is somewhat similar to that of the
Reissner-Nordstr\"{o}m black holes in that a countable number of
asymptotically flat universes are connected through compact regions
of inner and outer horizons as well as timelike singularities
(for more detail,  see  reference
\BSTRING), but the analogy should not be taken too seriously
since the nature of the inner horizons is very different.
The Penrose diagram of \metth\ cannot be accurately drawn in the
$r$-$t$ plane.

  For small $a$, a big mass gap of order $1/a$
develops for the internal degrees of
freedom and all the  low energy states have circularly
symmetric wave functions. For the most of this paper we will consider
the case of $a$ comparable to the Planck length so that the only
available internal state is the ground state. In other words, we
will consider the circularly symmetric sector of the theory.
In terms of the classical physics of the low energy observers, it means
the only relevant worldlines are those without any angular momentum.
In this section we want to explore the static geometry above as
seen by such  observers. To begin with, consider all geodesics
of vanishing angular momenta. This is consistent with the
geodesic equations of motion since $\theta$
is a Killing coordinate. With the help of two Killing vector fields,
the geodesic equations of motion can be reduced to the following form for
the $r$-coordinate as a function of an affine parameter $\tau$ \BSTRING.

\eqn\rgeo{({{\dot r} \over {2\lambda r}})^2=
         (1-{q \over r})(\epsilon^2+\alpha(1-{m \over r}))}

\noindent
$\epsilon$ is the covariant $t$-component of the 3-velocity
and $\alpha=-1,0,1$ for timelike, null and spacelike geodesics.
Immediately one finds a few revealing properties of $r(\tau)$.
For generic values of the energy $\epsilon$, $r$ bounces at $q$
quadratically, so that an initial inequality $r \geq q$ is
maintained throughout the history of the worldline. The only
exception occurs for spacelike geodesics with $1+{\epsilon}^{2}
=m/q$, in which case it takes infinite amount of time $\tau$
for $r(\tau)$ to reach $q$. Another important fact is that, for
$\alpha \neq 1$, $r=q$ is the unique extremal value inside
$r=m$. In other words, the time coordiate $r$ is monotonically increasing
toward both future and past of $r=q$, while $r<m$. Obviously,
no radial geodesics can penetrate the inner horizon and see the
curvature singularity at $r=0$ and the way it works out for
timelike or null worldlines is that all radial observers bounce
at the inner horizon into the next asymptotic region. In any case,
one can conclude the restricted region defined by $r \geq q$ is by itself
geodesically complete as far as radial geodesics are concerned.
The singular structure is completely hidden inside a barrier at $r=q$.
Therefore, the radial part of the metric \metth\ describes a nonsingular
2-D manifold with a countable number of universes as illustrated in
figure 1.

\ifig\Pen{Penrose diagrams of $g^{(2)}$ for various parameter ranges.
The bold lines indicate asymptotically far regions with vanishing
effective coupling $e^{\chi}$. $e^{\chi}=\infty$ at $r=q$.
$I^{+}$ and $I^{-}$ are the future
and the past null infinities of a particular universe}
{\epsfysize=6.0in\epsfbox{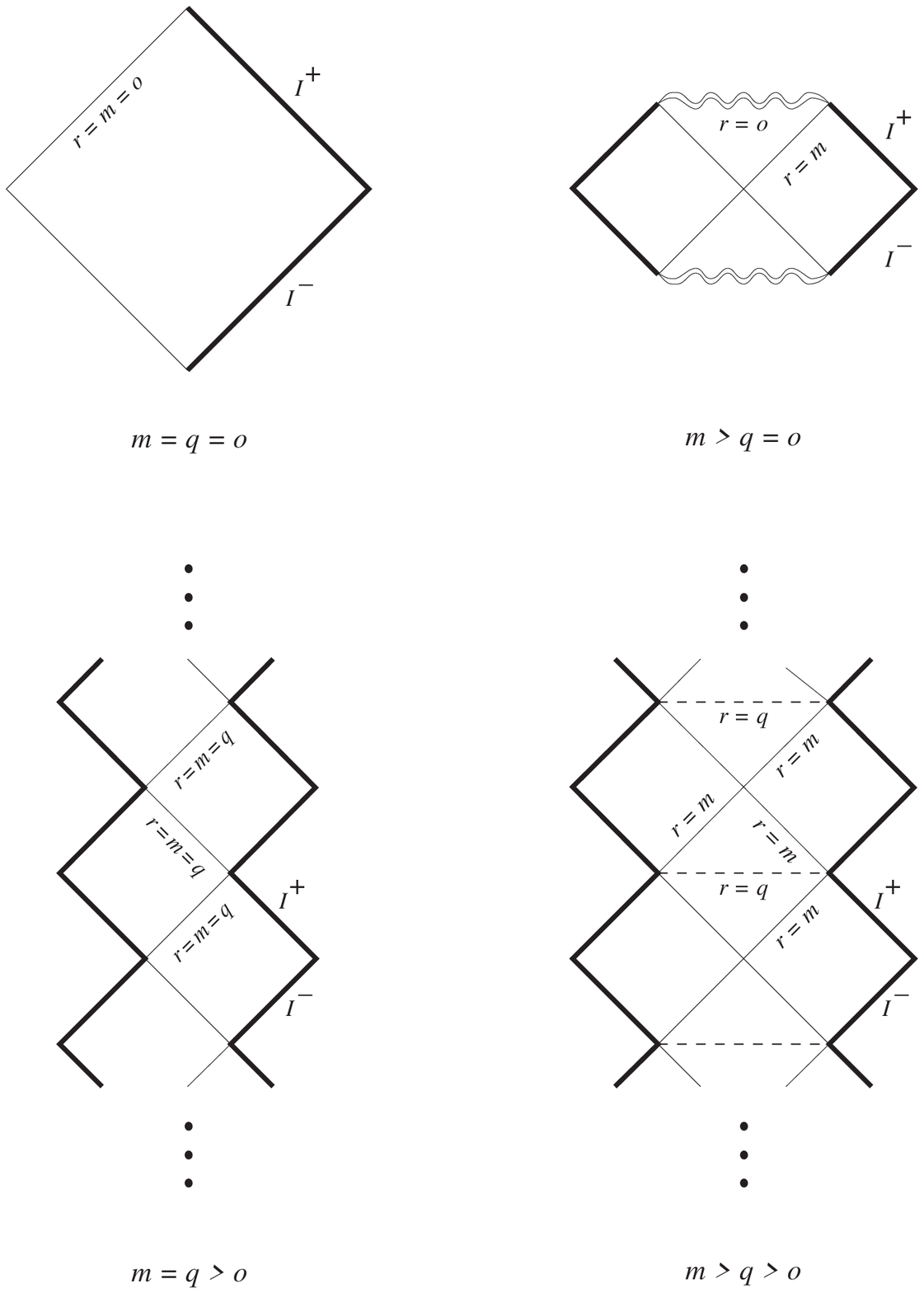}}

This unusual behaviour can be partially traced to the fact
$\theta$ becomes a time coordinate inside the inner horizon, where
a radially moving observer, freely falling or accelerated,
can never satisfy the on-shell condition.
Once we understand the peculiar nature of the inner horizons at $r=q$, it
is not difficult to draw the Penrose diagram of the 2-D metric, or
\metth\ without the $d\theta^2$ term, on the $r$-$t$ plane.
The nonextremal case is similar to that of the Reissner-Nordstr\"{o}m
geometry \ISRAEL\ with parts of it containing
the inner horizons and the singularities cut out and the remaining pieces
glued along spacelike hypersurfaces.
For this case, the inner horizons at $r=q$ become spacelike lines, to be
also called {\it the critical lines} from now on, contained in the
compact region surrounded by outer horizons $r=m$.
In figure 1, the
dotted lines of the case $m>q>0$ are these critical lines.
For $m=q=0$ they are null lines infinitely far away.
In the extremal limit $q \rightarrow m>0$, half of the universes are
decoupled and each remaining universe is successively attached
to the next by the outer horizons at $r=m$.
On the other hand the charge-zero limit can be easily shown to be
Witten's black hole with the spacelike singularities
forming along the critical lines (or the inner horizons) at $r=q$.
The change of variable, $r/\lambda=e^{2\lambda x}$, is necessary to go
back to a more familiar coordinate in  which the dilaton $\phi$
is linear.

   The action and the field equations governing the dynamics of
these 2-D spacetimes are obtained by a dimensional reduction of 3-D
effective string field theory \actth. Choosing a convenient set of
2-D fields and splitting the metric to $g^{(2)}$ plus the internal
part we obtain the following effective action.

\eqn\grav{S^{(2)}=\int \sqrt{-g^{(2)}} e^{-2\chi} (R^{(2)}+ 4
          (\nabla\chi)^2+4\lambda^2-(\nabla f)^2-e^{2f} K^2)}
$${g^{(3)}\equiv g^{(2)}+e^{-2f}a^2\,d\theta^2}$$
$${e^{-2\chi}\equiv e^{-2\phi-f}}$$
$${K_{ij}\equiv{1 \over 2a}H_{ij\theta}}.$$
\noindent
Here $K$ is an exterior derivative of a 1-form, hence a $U(1)$ gauge
field strength. The field equations are, after appending charged source on the
right-hand sides, the following: $Q$ is the total charge inside the
given value of $r$ and is locally constant in the source-free region.

\eqn\CHI{R^{(2)}+4\nabla^2\chi-4(\nabla\chi)^2+4\lambda^2
         -(\nabla f)^2+2Q^2e^{4\phi}=0}
\eqn\FFF{\nabla(e^{-2\chi}\nabla f)+2Q^2e^{4\phi-2\chi}=0}
\eqn\GIJ{-\nabla_{i}\nabla_{j}e^{-2\chi}+g_{ij}\nabla^2e^{-2\chi}
        +\cdots=T^{matter}_{ij}.}

\noindent
The 3-D field $\phi=\chi-f/2$ is kept for later conveniences and
$\cdots$ of \GIJ\ indicates terms of lower derivatives.
The Bianchi identity of the left-hand side of \GIJ\ implies the
energy momentum conservation of $T^{matter}$. It is not too difficult
to see that the static solutions of finite mass are all given
by the 2-parameter family above. For $q>0$ we can define
a new coordinate $y$ by $r=q \cosh^2{\lambda y}$, which shall be
useful later on.

\eqn\met{g^{(2)}=-(1-{m\over r})dt^2+{1\over 4\lambda^2\,(r-m)(r-q)}
          dr^2=-F\,dt^2+{1 \over F}dy^2}
$$F\equiv (1-{m \over q\cosh^2{\lambda y}})$$
$${e^{-2\chi}={r\over \lambda}\sqrt{1-{q\over r}}
={q\over \lambda}\cosh{\lambda y}\sinh{\lambda y}
\qquad e^{-f}=\sqrt{1-{q\over r}}=\tanh{\lambda y}}$$
\eqn\staR{R^{(2)}=4\lambda me^{2\phi}-6mqe^{4\phi}.}
\noindent
Notice the inequality $e^{-2\phi}\equiv e^{-2\chi+f} \geq {q/\lambda}$.
Even though the effective coupling $e^{\chi}$ is unbounded, the string
coupling $e^{\phi}$ is bounded for charged cases.
In the charge-zero limit, $f \equiv 0$, and
$\chi\equiv\phi$ becomes the usual dilaton of \CGHS.
The Hawking temperature \HAWK\ can be easily computed by requiring
the Euclidean version of $g^{(2)}$ to be nonsingular at the horizon
$r=m$. The periodicity of the Euclidean time coordinate is the inverse
temperature.

\eqn\hawktem{T_{Hawking}={\lambda \over 2\pi}\sqrt{m-q \over m}}
\noindent
\hawktem\ reduces to the expected value $\lambda/(2\pi)$
\CGHS\ as $q\rightarrow 0$.

To illustrate the relationship between these new spacetimes and
Witten's, recall the construction of the charged black strings \BSTRING.
A neutral black string is a simple product of a Witten's black hole
with a line. After a Lorentz boost along the line, one can make a duality
transformation \DUAL\ to get a new classical solution to the effective
string field theory with torsion, or charge as we call it in this paper.
We wrote the solution in \metth\ for a circle instead of a line.
Without any boost ($q/m=0$), the duality transformation is trivial and
$g^{(2)}$ is the original Witten's black hole. The maximal boost
corresponds to the extremal case $q/m=1$. If the Witten's black hole we
started with were an exact string background, the duality symmetry
of string theory would imply all nonextremal solutions of \met\ support
one and the same string theory. But as we shall see in the final section
this is not the case.

Simple counting reveals that this gravity system possesses one local
degree of freedom. Classically, this  means that there are numerous
time dependent matter free solutions. This poses an essential
difficulty in studying the dynamics and the question of stability
becomes quite nontrivial. In particular, one needs to solve the full
partial differential equations to study the process of gravitational
collapses. For a system devoid of a local degree of freedom,
such as the S-wave sector of Einstein-Maxwell theory or the dilaton
gravity of \CGHS, the solutions
are locally static wherever the matter is absent and for thin shells
of collapsing matter we can simply glue different static geometries
across the history of the shell \ISRAEL. This statement is usually
refered to as Birkhoff's theorem in general relativity.
But given a local degree of freedom,
the geometry will be fluctuating even after the matter part dies out
and can be found only by explicitly integrating the nonlinear
dynamical equations. For this reason we will investigate the
gravitational collapse in a piecewise manner and try to come up with
a physically reasonable scenario in sections 3 and 4.

\newsec{Dynamics of a Collapsing Massless Shell}

Discussions above raise an important question. Does the gravitational
collapse of charged matter leave behind a nonsingular spacetime
similar to the charged static solutions? A related question is whether
the inflow of charged matter to a Witten's black hole will lift the
curvature singularity just as the inflow of neutral matter in the linear
dilaton vacuum creates a singularity. An even  more immediate problem
is whether the nonsingular nature of the charged spacetime is stable against
extra (charged or neutral) matter inflow. A gravitational collapse
can be divided into the following three stages.

\noindent
(1) The immediate response of the geometry to the matter.

\noindent
(2) The residual gravitational fluctuation for $(t+\int{F^{-1}dy}) < \infty$.

\noindent
(3) The residual gravitational fluctuation as $(t+\int{F^{-1}dy})
                                               \rightarrow \infty$.

\noindent
While the general treatment of even small gravitational perturbation
is a fairly complex problem, the physics of (3) is relatively well
understood and is the subject of the next section. Such an asymptotic
tail of the gravitational fluctuation is known to be responsible
for the instability of the Cauchy horizon in the case of charged
black holes of the Einstein gravity.

One needs to analyze the effect of (2) to investigate the stability of
the event horizons, along the lines of Chandrasekar's analysis of the Kerr
black holes\ref\CHAN{S. Chandrasekar, The mathematical theory of black
holes, Oxford University Press, 1983.}. But knowing that the 2-parameter
family of solution \met\ is  the only finite mass static solution of
the theory and that the  ADM mass and
the charge are conserved quantities, it is difficult to imagine a possible
cause of instability that could change the structure outside the
event horizons.
Our static spacetime, however, has another source
of instability under (2), namely the infinite effective coupling at
the critical lines, which are the  inner horizons of the 3-D metric.
We will not pursue the matter here other than what is needed in
this section concerning certain gravitational shock waves. The difficulty
lies in  reducing the linearized equations (3 dynamical and 2
constraint) to a single unconstrained dynamical equation in the
region of interest. In other
words we were not able to solve the constraint equations.

   For part (1), consider a thin shell of conformal matter $T^{matter}$
collapsing in an initially static spacetime specified by the two
numbers $(m_s,q_s)$. (It is important to assume that the spacetime is static
initially in that we will need the explicit solution before the collapse.)
In the limit of infinitesimal thickness, that is for a shock wave,
the matter will induce discontinuous changes in normal\foot{\rm
Unlike the case of timelike shell, {\it normal} here does not mean
orthogonal. Using a null coordinate pair $(x^{+},x^{-})$, the normal
derivative is $\partial_{+}$ for a shock propagating along a fixed
value of $x^{+}$.} derivatives of various fields across the
shock. Furthermore one can obtain a set of first order differential
equations obeyed by these jumps from the set of the nonlinear field
equations of section 2. All one has to do is to take the discontinuities
of the homogeneous (source free) equations and to integrate the
inhomogeneous equations across the shock.

  Since we are dealing with massless matter, it is convenient to
introduce a light-cone coordinate in the conformal guage.

\eqn\conmet{g^{(2)}=-e^{2\rho}\,dx^{+}dx^{-}}

\noindent
Then $R^{(2)}=8e^{-2\rho}{\partial_{+}}{\partial_{-}}\rho$  and
${\nabla^2}=-4e^{-2\rho}{\partial_{+}}{\partial_{-}}$. Let
$x^{-}=-\infty$ on  past null infinity, to be concrete.
 Let the shock be centered at $x^{+}=x^{+}_{o}$.  The only nonvanishing
component of the matter energy momentum tensor is $T^{matter}_{++}$, which
has an infinitesimally thin support.  Then we can assume that
$\chi$ and $f$ are continuous across the shock while the
derivatives along $x^{+}$ may not be. Furthermore we assume the continuity
of $\rho$ as well, the  consistency of which needs to be checked.
First of all the effect of $T^{matter}_{++}$ is felt by the dilaton $\chi$
through the $\delta g^{++}$ equation of \GIJ. If we define $\sigma$ as
the total flux, we have the equation

\eqn\defSIG{\sigma(x^{-})\equiv \lim_{\delta \rightarrow 0}
           \int^{x^{+}_{o}+\delta}_{x^{+}_{o}-\delta} T^{matter}_{++}
           dx^{+}=\lim\int -\nabla_{+}\nabla_{+}e^{-2\chi}+\cdots
           =-[\partial_{+}e^{-2\chi}].}

\noindent
Here, we introduced the notation [A] for the difference of A across the shock.
Next, the evolution of $\sigma$ is governed by $\delta g^{+-}$ equation,
whose difference across the shock is

$${\partial_{-}\sigma=-\partial_{-}[\partial_{+}e^{-2\chi}]
          =-e^{4\phi_{o}+2\rho_{o}-2\chi_{o}}{[Q^2] \over 2}.}$$

\noindent
Here $\chi_{o}$ denotes $\chi$ restricted to $x^{+}=x^{+}_{o}$ and
similar for the other fields. Note that we already know values of these
$\chi_{o},\cdots$. They are simply the corresponding static values
owing to the continuity. Naturally all the quantities above are
functions of $x^{-}$ only. On the other hand the energy momentum
conservation of $T^{matter}$ implies

\eqn\mass{\partial_{-}\sigma=-[T^{matter}_{+-}]=0.}

\noindent
Obviously the procedure is inconsistent for a charged shell
($[Q^2]\neq 0$). The assumption of continuous $\rho$,
needed when we derived \defSIG\ ignoring the
connection piece $\Gamma^{+}_{++}\partial_{+}e^{-2\chi}$, is apparently
too strong. A priori, there is no reason that $\rho$, a coordinate
dependent function, should be continuous. However for the case
of {\it massless and neutral\/} shell ($[Q^2]=0$), we shall find
no further inconsistency below. For this reason we will consider
neutral conformal matter only, so that $\sigma$ is constant.

We can also take the discontinuity of the equations
\FFF\ and \CHI\ and the result can be written in the form (after
substituting $\sigma$ for $-[\partial_{+}e^{-2\chi}]$)

\eqn\delf{\partial_{-}[\partial_{+}f]-\partial_{-}\chi_{o}[\partial_{+}f]
         -{e^{2\chi_{o}} \over 2}\sigma \partial_{-}f_{o}=0,}
\eqn\crvt{{e^{-2\chi_{o}+2\rho_{o}}\over 8}[R^{(2)}]
           -\sigma\partial_{-}\chi_{o}
         +{e^{-2\chi_{o}}\over 2}\partial_{-}f_{o}[\partial_{+}f]=0.}

\noindent
Reduced to the usual dilaton gravity ($f\equiv 0,Q\equiv 0$), with the
Kruskal coordinate in which $e^{-2\chi_{o}}=e^{-2\rho_{o}}=m/\lambda-
\lambda^2 x^{+}_{o} x^{-}$, these equations have the solution

$$[R^{(2)}]=4\lambda\sigma x^{+}_{o} e^{2\chi_{o}}.$$

\noindent
$e^{2\chi_{o}}=e^{2\phi_{o}}$ is unbounded and shows a curvature singularity
forming at infinite effective coupling $e^{\chi}=\infty$. It is easy to
see the result reproduce what we would find by actually solving the
full equations as done in  \CGHS.
In this particular case there is no gravitational
fluctuation afterwards and the solution outside the shock is completely
determined by this jump. All we have to do is to find the right static
solution to match.  It is worthwhile to notice that, in spite of
the singular behaviour, $\sigma$ is constant and therefore $e^{-2\chi}$
is continuous even at the singularity, a neccessary behaviour for
the self-consistency.

Now what happens if we start with a charged spacetime? Then we have
to solve for $[\partial_{+}f]$ first. Again for the sake of consistency
we will consider a shock of neutral matter in an initially charged
spacetime. Solving \delf,

\eqn\delfone{e^{-\chi_{o}}[\partial_{+}f]=(C+{\sigma \over 2}
           \int^{x^{-}}_{-\infty} e^{\chi_{o}}\partial_{-}f_{o}\,dx^{-}).}

\noindent
Since the matter shock is neutral $\sigma$ is independent of $x^{-}$.
$C$ is an integration constant. The lower bound of the integral is past
null infinity where the shock wave originates.

Consider the effect of $C$ on the jump of the curvature.
Near the critical line $e^{-2\chi}=0$ (or $y=0$ in the coordinate of \met),
the corresponding jump scales like $\sim C/\sqrt{y^3}$ and is
infinite at $y=0$.
This is rather strange in that not only the strength of the curvature
singularity but also its sign are completely independent of
the matter you are throwing in. (As to be shown later, the contribution
from $\sigma$ is finite at the critical line.) Even more disturbing
is the jump of the internal part of the Ricci tensor $R_{\theta\theta}
=-e^{f}\nabla^2 e^{-f}$.  Asymptotically it is $\sim e^{-\lambda y}$,
whereas the static value is $\sim e^{-2\lambda y}$. This seems to suggest
that the gravitational fluctuation with nonzero $C$ would cost an
infinite amount of energy. It turns out that the problem is closely
related to the asymptotic nature of the gravitational perturbation.

As demonstrated in the appendix, the gravitational perturbation around
a static solution is asymptotically described by the following massive
Klein-Gordon equation with $\Psi \equiv e^{\chi_{s}}(f-f_{s})$.

\eqn\KG{\nabla^{2}\Psi-\lambda^2\Psi=0.}
\noindent
$\chi_{s}$ and $f_{s}$ are  the static values of $\chi$ and $f$.
Across any null line $x^{+}=x^{+}_{o}$, $[\partial_{+}\Psi]=C$ is
allowed by \KG. But it is quite deceptive since the behaviour  of
field as we move away from the null line is profoundly affected
by the massive nature of the evolution equation. After all, we do not
expect a massive field to propagate along a null line.
As shown in the appendix, the bump of $\Psi$ itself owing to the
discontinuous derivative tends to diminish rapidly with time and
becomes completely undetectible after an infinite amount of time.
In other words, if we incorporate such a jump to the initial condition
on past null infinity, the field configuration of $\Psi$ in finite
region does not show any sign of discontinuous derivatives. (In terms of the
homogeneous solution presented in the appendix, this case corresponds to
$u_{o} \rightarrow -\infty$ before taking the normal derivative.)
On the other hand, if we insist that the field configuration show
such a jump away from past null infinity,  the required initial condition
on past null infinity has divergent $\Psi$, and the energy content of the
initial flux of the associated gravitational radiation
is infinite. It is simply impossible to produce  nonzero effective
$C$ in the finite region with a finite amount of energy.\foot{\rm
See appendix for detailed discussion on the asymptotics of the
gravitational perturbation.}  $C$ should be dropped.

While we discussed the problem in the context of a gravitational
collapse of massless matter fields, the conclusion must hold in other
contexts. After all, $C$ represents a homogeneous
solution to the gravitational field equations. We should not think
of it as being generated by the matter shock. If $C \neq 0$ were
allowed, it would imply the instability of the critical line
under gravitational perturbation automatically, for example. In
this sense we have eliminated one possible way for the infinite
self-coupling to cause instability.

Rewriting \delfone,

\eqn\delftwo{[\partial_{+}f]={\sigma\over 2} e^{\chi_{o}}(I(x^{-})-I(-\infty))
           ={\sigma\over 2} e^{\chi_{o}} I(x^{-}).}

\noindent
$I$ is the indefinite integral of $A(x^{-})\equiv e^{\chi_{o}}\partial_{-}
f_{o}$  over $x^{-}$ and invariant under coordinate tranformations $x^{-}
\rightarrow z^{-}(x^{-})$. One can ask whether the massive nature of the
graviton affects this inhomogeneous part of the solution. The answer is no.
It turns out that the asymptotic behaviour of the {\it source} term
$A$ is such that a separation of variables occurs. The  form of $\Psi$ near
the shock is given by a product of two factors each of which
is function of one of the null coordinates only. As a result, the estimate
\delftwo\ is reliable even though we derived \delf\ without taking into
account the massive nature of the gravitational perturbation.
Inserting \delftwo\ to \crvt,


\eqn\mainR{[R^{(2)}]=2\sigma e^{-2\rho_{o}} (2\partial_{-}e^{2\chi_{o}}
           -e^{\chi_{o}} \partial_{-}f_{o}I).}

\noindent
No infinite jump is possible except maybe at the critical point
$e^{\chi_{o}}=\infty$. But even at this point the leading terms of two
parts cancel each other and the true leading term is a constant.
For  an initially nonextremal spacetime, an easy way to see this is
to make a coordinate transformation $z^{-}\equiv y(x^{+}_{o}, x^{-})$
where $y$ is the coordinate used in \met\ and recall that
$e^{-2\chi_{o}} = qy+O(y^3)$, $e^{-f_{o}}=y+O(y^3)$.

\eqn\finiteR{-\infty < R^{(2)} < \infty
\qquad\hbox{near}\quad x^{+}=x^{+}_{o}.}

\noindent
The conclusion holds for all initially charged cases including
the extremal case.
Therefore the history of a collapsing shell of neutral
conformal matter is qualitatively
similar to that of a massless observer. It just bounces into the
next universe. It is quite a surprising behaviour when compared to
what happens in the usual dilaton gravity \CGHS.

\ifig\Cauchy{A collapsing shell of neutral conformal matter
in a nonextremal spacetime.
The shell simply bounces into the {\it next universe}.
The wiggly arrows indicate the residual gravitational fluctuations,
some of which enter the future event horizon (EH) at arbitrarily
late times and propagate parallel to the Cauchy horizon (CH). Again,
the dotted line is the critical line of infinite effective coupling.}
{\epsfysize=4.5in\epsfbox{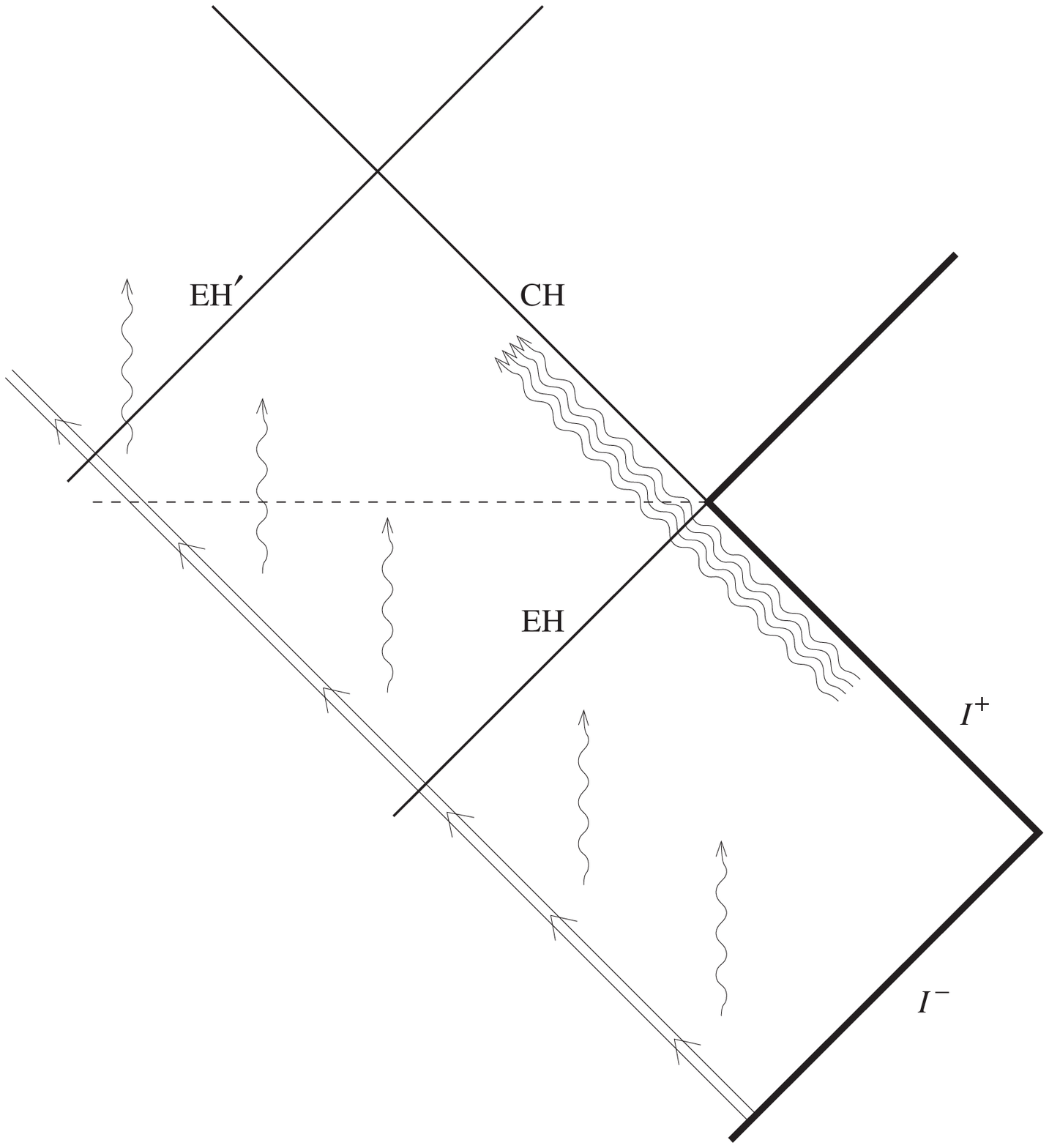}}

 To conclude this section consider the case of an initially
neutral spacetime. Will the singularity be lifted by the inflow
of some charged matter? Can we open up a curvature singularity by
a completely classical process? As we have seen earlier, we should be
more careful in dealing with charged matter. In principle, one may
proceed by considering two conformal factors inside and outside the shock,
each of which is smooth but assumes values different from the other
along the shock.
However, we will postpone this problem of the gravitational collapse of
charged matter in a neutral spacetime to a future project for the following
reason. A neutral static spacetime has $e^{-2f}\equiv 1$, while the charged
case has $e^{-2f}=\tanh^2{\lambda y}$ ranging from $1$ to $0$.
In the framework we are using, $e^{-2f}$ is a scalar and should be
continuous across the shock. Then the resulting immediate response of the
initially neutral solution to the collapsing shell of charged matter
must have constant $e^{-2f}$ and hardly resembles a static charged
spacetime. Even though it might be that the solution settles down
to a nonsingular spacetime with varying $e^{-2f}$ after a while,
we can not detect that with this type of local analysis. At the singularity
the scalar field can actually jump discontinuously to the desired value,
but such a behaviour could imply an inconsistency. The same argument
applies to the original 3-D dilaton field $e^{-2\phi}$.
In this sense the shock wave anaylsis seems less promising for
initially neutral cases. Even if we carried out similar calculations for the
case of a charged shell, we would be unable to decide whether the spacelike
singularity was lifted. Probably we  need to go beyond the immediate
neighborhood of the shock to probe the geometry outside the shock.

A different way of tackling this problem is to try to establish
the stability of the critical lines, which would certainly imply
{\it time dependent\/} charged solutions without a spacelike singularity.
In this regard,
the results above on the neutral matter shock in an initially charged
spacetime are quite encouraging. We found that, as a consequence of the full
nonlinear dynamics, no singularity met the shock.
If the critical lines are unstable, it is difficult to imagine
where the resulting spacelike singularity along the critical line might start.
It has to start somewhere away from the shock.

\newsec{Instability of the Cauchy Horizon}

In section 3, we found that a collapsing shell in an initially
charged static spacetime does not encounter a curvature singularity
and bounces into the next universe. However, any gravitational
collapse is followed by residual gravitational fluctuations and
so far we have not addressed the effect of them on the subsequent
evolution of the geometry. In principle these fluctuations could
grow indefinitely to produce curvature singularities. The most
obvious place to look for the singularity is the Cauchy horizons,
which are generically unstable
because of the infinite blue shift of the asymptotic residual
fluctuation entering the future event horizon at arbitrary late times.
However, unlike the more familiar case of Reissner-Nordstr\"{o}m black holes
the Cauchy horizons of our geometry coincide with the outer horizons
rather than the inner horizons. The inner horizons of the nonextremal
geometry, also called the critical lines, are effectively
spacelike for any radial observer and cannot be the Cauchy horizons.
While the behaviour of the critical lines
under the gravitational perturbation (as opposed to the matter inflow) still
needs to be studied, we will first concentrate on the more obvious
potential source of instability.

 A convenient way to analyze the instabilty of the Cauchy horizon is
to consider ingoing null flux of energy stretched all the
way to the infinite future \POISSON.
Assuming an initially static geometry \met,
choose a new set of coordinates $(v,u)$.

\eqn\newmet{g^{(2)}= -F(y)\,dt^2+{1 \over F(y)}dy^2
                   = -F(y)\,dvdu
                   = -F(y)\,dv^2+2\,dvdy}
$${dv=dt+dy/F,\quad du=dt-dy/F}.$$

\noindent
With this coordinate system  the Cauchy horizon of our universe is at
$v=+\infty$ inside the future event horizon. In
figure 2, we labeled the Cauchy horizon by CH and the future event
horizon by EH. Outside EH, $v=+\infty$ corresponds to future
null infinity $I^{+}$ and we are interested in asymptotic
inflow of the radiation stretched all the way to $I^{+}$.

For simplicity, suppose that the energy inflow is due to matter
rather than  gravitational radiation.\foot{\rm
We are indebted to E. Poisson for suggesting this approach.}
This simplification is well justified by the universal nature
of the instability.
By the energy-momentum conservation, the
energy-momentum tensor of a massless pure inflow is of the form

\eqn\emtens{T^{matter}= \mu(v)\,dvdv.}

\noindent
As $v \rightarrow \infty$ (future null infinity if outside the event
horizon) a typical $\mu(v)$ shows an inverse
powerlike behaviour \PRICE\ with  finite total energy and has little
effects on the geometry outside the event horizon. But inside the event
horizon, $v=\infty$ corresponds to CH at finite physical distance and
the energy  density measured by a freely falling observer could be
infinite  at CH. To be definite  let us choose
Kruskal type coordinates $(V,U)$, to be defined later,
good for both CH and EH. In addition, let $V=0$ at CH,

\eqn\Kruskal{g^{(2)}=-e^{2\rho}\,dVdU\qquad (V=V(v),U=U(u)),}
\eqn\CHDiv{T^{matter}=\mu(v)\,dvdv=\mu(v)\,({dv/dV})^2\, dVdV
           \equiv T_{VV}\,dVdV.}

\noindent
Notice that once the flux crosses future event horizon (EH),
it looks more like
a thin shell of massless matter, even though $\mu(v)$ has an
infinite span outside. The energy density measured by a timelike
observer crossing CH is roughly $\sim T_{VV}$ which
diverges unless the fall-off of $\mu(v)$ is fast enough to overcome
the infinite blueshift factor $({dv/dV})^2 \rightarrow \infty$.

A better estimate can be made using the machinery of the previous
section. Since only the large $v$ behaviour is important, let
$\mu$ be nonzero only for $v>v_{min}$ with sufficiently big $v_{min}$
, so that near CH the flux is that of an arbitrarily thin shell.
Then certainly the formalism of section 3 is applicable inside
EH.
Using the definition \defSIG, after identifying the corresponding
null coordinates $x^{+}\equiv V$ and  $x^{-}\equiv U$,

\eqn\totmass{\sigma=\int^{0}_{V(v_{min})}T_{VV}\,dV
             =\int^{\infty}_{v_{min}}\mu(v)\,({dv/dV})\,dv .}
\noindent
As long as $\sigma$ is finite, we can estimate the jump of the
curvature on CH induced by the neutral inflow using \crvt.
For initially static spacetimes,
$${[R^{(2)}]_{V=0}=0.}$$

\noindent
The reason is simply that the dilatons
$\chi_{s}$ and $f_{s}$ are uniform along CH and
$\partial_{U}\chi_{o}=\partial_{U}f_{o}=0$.
This is a somewhat fictitious situation
since we have assumed that  before the
on-set of the asymptotic inflow at $v=v_{min}$ the geometry is
strictly static. For a realistic model, the asymptotic inflow
originates from the collapsing matter and the geometry is
non-static even for $v<v_{min}$. In particular CH does not
neccessarily coincide with an apparent horizon, the dilatons $\chi$,
$f$ are no longer uniform along CH, and as a result the
curvature jump is  generically nonzero and finite, proportional to $\sigma$,

$$[R^{(2)}]_{V=0} \sim \sigma \quad\hbox{up to some finite factor
as a function of }U.$$

\noindent
A similar mechanism must work for infinite $\sigma$, even though the
formalism of section 3  does not apply in this case. After taking
into account other residual fluctuations, an infinite $\sigma$
would most probably generate a null singularity along CH.
The role of the extra radiation flowing into the Cauchy horizon
rather than parallel to it has been first studied in \POISSON\
and found crucial for the instability to manifest itself.

For the particular case of the compactified charged black strings we
are studying, we can choose the following Kruskal coordinates

\eqn\nonS{V_{nonextremal}=-{1 \over \beta}e^{-\beta v} \quad \beta=\lambda
          \sqrt{m-q \over m},}
\eqn\exS{V_{extremal}=-{\lambda^2 \over v}.}

\noindent
(Obviously, this choice is not unique, but the same conclusion holds
for any other  coordinate regular on CH.)
For any powerlike $\mu(v)$, $\sigma_{nonextremal}$ is infinite, and the
Cauchy horizon is inherently unstable. Unless the asymptotic residual
fluctuation turns out to be exponentially small ($\sim e^{-\lambda v}$
for example), a null curvature singularity will form along CH.
But provided that $\mu(v) \sim
v^{-n}$ with $n>3$, $\sigma_{extremal}$ could be finite.
Whether a finite $\sigma_{extremal}$ indicates a stable Cauchy horizon of an
initially extremal solution is unclear. The extremal geometry can
always be changed to a nonextremal one by an infinitesimal increase of the
mass, and once that happens we must use the nonextremal estimate of
$\sigma$.

(However, this classical picture might undergo a drastic modification
once we include the semiclassical effect of Hawking radiation. Since the
temperature is positive except for the extremal spacetime, the geometry
outside the event horizon must approach the extremal geometry in the
infinite future ($v\rightarrow \infty$). If the same is true for the
geometry inside the event horizon, we must use the extremal estimate of
$\sigma$ and the Cauchy horizon might as well be stable for any initially
charged geometry.)

What are the consequences of the null singularity along CH?
Note that the Cauchy horizon of our universe (CH) meets
the future event horizon of the next universe. But
all the infalling observers escape to the next universe
through its past horizon (EH'). That is, none of the
observers from our universe, entering EH before the on-set of the
asymptotic inflow \emtens, cross CH between EH and EH'.
The singularity on CH hardly interferes with the observers
entering the next universe. While some  massive observers
without sufficient momenta would be attracted to the singularity, the rest
will probably continue to the asymptotic region away from it.
This should be compared to what happens in case of 4-D
Reissner-Nordstr\"{o}m (nonextremal) black holes. For these 4-D
black holes the phenomenon is well known as {\it{mass inflation}\/}
\POISSON. The static geometries of nonextremal Reissner-Nordstr\"{o}m
black holes allow timelike observers (entering the future event horizon)
to escape to the next universe,
the static singularity being timelike. Once we include generic
perturbations, however, the Cauchy horizon becomes singular
all the way to the point where it meets the original timelike singularity
and even the timelike observers  eventually experience infinite tidal
force.\foot{\rm However, it has been shown the total impulse from
the tidal force is finite and the would-be Cauchy horizon
is transversible \ORI.}

Of course, the past event horizon of the next universe (EH') overlaps
the Cauchy horizon of another universe which is spacelikely
separated from our universe. An asymptotic inflow from that universe
can destabilize EH'. But if we suppose the  charged spacetime is made by
a gravitational collapse in our universe, the extended static
structure before the collapse is
fictitious just as the past event horizon of the Schwarzschild
geometry is fictitious for a collapsing star. Then EH', the past event
horizon of the next universe, would be as stable as EH.

\newsec{Discussion: Nonsingular Exact String Backgrounds}

The most curious feature of the charged  black string is probably
the way the inner horizons conceal the curvature singularities from
the radial observers. The resulting 2-D geometry is geodesically
complete and the scalar curvature is finite everywhere.

\eqn\boundR{-\lambda^2{2m\over q} \leq R^{(2)} \leq
           \lambda^2{2m\over 3q}.}

\noindent
In particular, as $q \rightarrow m$, the absolute value of the
curvature is bounded by $2\lambda^2$. On the other hand, for Witten's
black hole seen as the dimensional reduction of the 4-D dilatonic
black hole, $\lambda^{-1}$ corresponds to the total magnetic charge \GIDD,
or the radius of the long necked extremal solution called
the cornucopion \REMNANT. If we take this parameter $\lambda$
much less than the Planck mass, then, as the black hole evaporation
progresses $m \rightarrow q$, the geometry is of macroscopic scale
everywhere including inside the event horizons. Furthermore the
string coupling $e^{\phi}$ is bounded by $\sqrt{\lambda /q}$.
It is possible that the quantum fluctuation of geometry does not
change the causal structure qualitatively.
Taken seriously, this last statement
has a profound ramification in the black hole physics: Whatever the
real answer is to the information puzzle\foot{\rm For a
recent review  see the reference \PRESK.}\HAWKINF,
it has little to do with the
details of the Planck scale physics. Either the information is
completely recovered before the collapsing matter enters the future
event horizon, or some will  be lost forever to the next universe.

Any remnant scenario is likely to assume some
nonsingular objects as the final
products of the evaporation process. They are assumed to be able to
carry macroscopic amounts of the information yet interact very little
with the surroundings. All these assumptions may or may not be realized
depending on the details of the Planck scale physics. Even in the
case of the cornucopions we still need to explain how the information
inside the horizon, far down at the tip of the long neck,
propagates out to occupy the whole length of the cornucopions, which
could happen only near the end of the process where the singular structure
inside the horizon must be resolved somehow. The
static extremal solution above could obviate all these details.

But we are hardly in a position to take these static solutions seriously.
The Reissner-Nordstr\"{o}m nonextremal solution, which is also known to
allow some of the infalling observers (massive ones) to escape to the next
universe, is  dynamically unstable  and the
resulting generic geometry has a curvature singularity
blocking the passage though perhaps not completely \POISSON\ORI.
In the present case of the compactified black strings,
while the instability of
the Cauchy horizon due to the asymptotic inflow from our universe
does not block the passage, there is a possibility that the
critical lines of the infinite effective coupling become singular
under the generic gravitational perturbation. As mentioned at the end
of section 3, the fact that a collapsing shell does not encounter
a singularity even at the critical line seems to suggest no singularity
whatsoever along the entire span of the critical line, but this
might be an artifact of the initially static geometry.

Another difficulty concerns the creation of the charged spacetimes
from vacua. If the linear dilaton background is the true vacuum, it is
very difficult to imagine how the extended structures of the charged
spacetime could be made from the gravitational collapse of thin shells
without any singularity at the infinite effective coupling.
More specifically, it is rather difficult to
draw a sensible Penrose diagram describing the process.  Even a
qualitative understanding in this regard would be an important
step.

At this point one can easily see that all the unanswered questions
are in one way or another associated with the true nature of
the critical lines (the inner horizons  as seen by 3-D observers),
which become curvature singularities under the duality transformation
(see section 2).
The ultimate question is then, {\it between the two geometrical
pictures dual to each other, namely the Witten's causal structure and
our nonsingular version, which one of them resembles the reality
more closely} \DUAL\BSTRING.

Of course the effective field theory \actth\ we started  with is correct
only up to $O(1)$ in the $\alpha'$ expansion of the sigma-model \SIGM.
That is, the extended nature of the fundamental string is not properly
taken into account. In particular, the static solutions we
have found are related to the coset models
$(SL(2,R)_{k}\otimes U(1))/U(1)$ for $m\geq q>0$,
or $(SL(2,R)_{k}/U(1))\otimes S^{1}$ for $q=0$, with
$(k-2)=(2\alpha'\lambda^2)^{-1}$. $k$ is larger than $2$ provided
that the total dimension of the spacetime, compact or not, is less
than the critical dimension.
A few months after Witten's derivation of the 2-D black hole
geometry from $SL(2,R)_{k}/U(1)$ coset model, R. Dijkgraaf et.al.,
attempted to find the exact geometry of the coset model by investigating
the tachyon spectrum of the theory \VVD\ with the assumption that the string
on-shell condition $L_{0}=1$ should be interpreted as the tachyon field
equation $\nabla^2 T=0$. Recently, A.A. Tseytlin rediscovered the same
metric in a functional integral approach where he replaced the
classical effective action of the coset WZW model by a quantum effective
action, which involved modifying the coefficients of the action from
$k$ to $k+c_{G}/2$ \TSEY. The change of metric can be easily written
in the following form, in a gauge similar to those used in \WITT\VVD\TSEY.

\eqn\CHange{g^{(2)}_{Witten}={1\over \lambda^2}dx^2-\tanh^2x\,dt^2
            \quad\rightarrow\quad g^{(2)}_{exact}={1\over \lambda^2}dx^2-
            {(1-2/k)\tanh^2 x\over 1-(2/k)\tanh^2 x}dt^2}
\noindent
In the gauge of \metth, this can be rewritten

\eqn\exactmet{g^{(2)}_{exact}=-(1-{m\over r})\,dt^2+{1\over 4\lambda^2
              (r-m)(r-(2/k)m)}dr^2.}

\noindent
Amazingly, this looks exactly like a nonextremal metric, \met\ with
$q/m\rightarrow 1>2/k>0$. The $SL(2,R)_{k}/U(1)$ coset metric has the causal
structure of the  Penrose diagram labeled by $m>q>0$ instead of the
singular one ($m>q=0$) in figure 1. Furthermore the dilaton part is,
according to Tseytlin's estimate,

\eqn\exactdil{e^{-2\chi_{exact}}={r\over \lambda}
             \sqrt{1-{(2/k)m\over r}}.}

\noindent
Again, it is identical to that of the nonextremal solution \met\ with
$q/m=2/k$. The implication is very remarkable. The reality seems to be better
described, at least classically, by the nonsingular causal structures
we have found than by Witten's.

Another encouraging evidence for these nonsingular solutions comes
from the duality symmetry of the string theory.
The $(SL(2,R)_{k}\otimes U(1))/U(1)$ coset model has been considered
by I. Bars and K. Sfetsos \ref\Bar{I. Bars and K. Sfetsos,
{\it Exact effective action and spacetime geometry in gauged WZW models},
USC-93/HEP-B1; K. Sfetsos, {\it Conformally exact results for $SL(2,R)\otimes
SO(1,1)^{(d-2)}/SO(1,1)$ coset models}, USC-92/HEP-S1.},\foot
{\rm We thank G. Horowitz for drawing our attention to this work.}
also nonperturbatively.
The resulting metric is singular, unfortunately, and the causal stucture
is again that of timelike singularities hidden behind inner and outer
horizons. However, if we consider the 2-D part of the metric as we
did in section 2,

\eqn\BSexact{g^{(2)}_{BS}=-(1-{m\over r})\,dt^2+
             {1\over 4\lambda^2(r-m)(r-q-(2/k)(m-q))}dr^2,}

\noindent
after an appropriate redefinition of the parameters in \Bar.
This metric is identical to the $O(1)$ approximation \met\ except that
$q\geq 0$  is replaced by $q+(2/k)(m-q)>0$.
The case of $q=0$ coincides with $g^{(2)}_{exact}$ as it should, and
the extremal limit is achieved by letting $q\rightarrow m$.
(We need to perform a coordinate rescaling on the metric presented
in \Bar\ to acheive this limit.) Now since the $O(1)$ approximations
to $g^{(2)}_{exact}$ and $g^{(2)}_{BS}$ are related to each other
by a duality transformation, it is not unreasonable to expect these
nonperturbative versions are dual to each other.

Unlike the $O(1)$ approximations, the 2-D part $g^{(2)}_{BS}$ of the
nonperturbative black string metric is  qualitatively self-dual.
The causal structure of the 2-D part does not change
upon $g_{\theta\theta}\rightarrow (1/g_{\theta\theta}),\cdots$,
with the exception of the extremal case.
We no longer have to choose between two
entirely different causal structures dual to each other.
It is true that there are many examples of different geometries (or even
different topologies) supporting the same string theory, and a fundamental
string must have a unique classical history independent of the
particular geometry  chosen. But here, instead of the horrendous task
of figuring out the actual behaviour of test strings inside the
event horizon, we are presented with a unique nonsingular causal
structure invariant under the duality transformation. The naive expectations
based on the behaviour of test particles are unambiguous and might as well
indicate the actual behaviour of test strings.
A radially moving classical fundamental string propagating into the
future event  horizon does not see a curvature singularity and simply
propagates  to another universe.

As for the exceptional case of the extremal solution, we can easily
see that it is dual to the linear dilaton background.
The relationship between masses of a pair of dual solutions is given by
(according to the $O(1)$ estimate)

$$m_{after}=m_{before} \cosh^2{\alpha},$$

\noindent
where $\alpha$ is the Lorentz boost parameter; $\alpha$ is infinite
for the maximal boost \BSTRING. Obviously a finite mass extremal solution
is obtained by the duality transformation on a maximally boosted
zero mass limit of $SL(2,R)_{k}/U(1)\otimes S^{1}$, the linear
dilaton background up to a trivial internal part. The positive mass of
the extremal solution must be an artifact of the effective field theory
estimate. While this ambiguity prevents us from viewing
either of them as the true ground state, it is quite interesting in that
we have an alternative description for the end stage of the black holes,
which has an event horizon at finite physical distance and another universe
beyond it.

Of course we overlooked two important facts in the arguments above.
First of all, the effective coupling is again unbounded.

\eqn\BSdil{e^{-2\phi_{BS}}={r\over \lambda}\sqrt{1-{(2/k)(m-q)\over r}},}
\eqn\BSeff{e^{-2\chi_{BS}}={r\over \lambda}\sqrt{1-{q+(2/k)(m-q)\over r}},}

\noindent
which we deduced from  \Bar. The effective coupling
$e^{\chi}$ is infinite at $r=q+(2/k)(m-q)$. In section 3,
we have seen this infinity was not strong enough to create a curvature
singularity out of an collpasing shell of matter. But we have not
been able to address the problem of gravitational perturbation completely,
let alone quantum fluctuations.  Secondly, the 3-D metrics of the
compactified charged
black strings ($m>q>0$),  perturbative or nonperturbative, possess
not only timelike singularities but also closed timelike worldlines
inside inner horizons. While it is reassuring that the corresponding regions
are completely inaccessible to  2-D observers and that they are
causally disconnected from us by event horizons, it is quite unclear
whether such an background is physically acceptable. It might be that
we should be content with the effective field theory \grav\ rather than
try to interpret the 2-D solution as a part of a classical string
background.

Finally, we would like to mention that similar causal structures
have been discussed in different
contexts in \GHOST\TRIVEDI\NEW. After the completion of this work,
we received a preprint by M.J. Perry et.al., \ref\Perry{M.J. Perry
and E. Tao, {\it Nonsingularity of the exact two-dimensional string
black hole}, DAMTP R93/1.}, which discussed the causal structure
of $g^{(2)}_{exact}$ above. They found the same causal structure as we did.
Also A. Tseytlin informed us of \ref\TV{A. Tseytlin and C. Vafa,
Nuc.Phys.B372 (1992) 443.} where he and C. Vafa observed that
the nonpertubative coset metric \CHange, of R. Dijkgraaf et.al., remained
bounded when continued to imaginary $x$.

It is my pleasure to thank J. Park, E. Poisson and K. Thorne for
sharing their expertise on general relativity. I would like also
to thank J.H. Schwarz, S. Trivedi and my advisor J.P. Preskill for the useful
conversations as  well as their kindly encouragement.

\appendix A{Asymptotics of the Effective Theory}

To study the asymptotic behaviour of the gravitational perturbation,
it is convenient to start with the ADM mass formula, obtained using
the canonical formalism  \ref{\TEITEL}{T. Regge and C. Teitelboim,
Ann.Phys.88 (1974) 286.}\ on \grav.
$\Delta$ denotes the deviation from the linear dilaton vacuum.

\eqn\ADM{M_{ADM}=2\,Ne^{-2\chi}\Delta({\chi' \over
         \sqrt{g}})\biggr{\vert}_{x \rightarrow \infty}}
$${g^{(2)}=N^2dt^2+g\,{(dx+L\,dt)}^2 \qquad N,g
           \rightarrow 1;\, L \rightarrow 0 \qquad\hbox{as}
           \quad x \rightarrow \infty}$$
\eqn\vacua{\chi_{vacuum}=-\lambda x \qquad f_{vacuum}= 0.}

\noindent
Readers should be warned that \ADM\ is derived with certain
assumption on the allowed phase space \TEITEL. More explicitly we need to
restrict the phase space to  $\Delta\chi_{s} \sim
\Delta \sqrt{g_{s}} \sim \Delta f_{s} \sim e^{-2\lambda x}$ at most,
which is sensible for the static solutions. Applied to \met\ we
obtain $M_{ADM}=m$.

Now consider a time-dependent solution. The requirement of finite mass
restricts the possible asymptotic behaviours \ADM. Obviously the same
asymptotic restrictions apply to $\delta \chi$, $\delta \rho$,
the corresponding deviations from a static solution. To be definite,
write a time dependent solution in the following form.

\eqn\permet{g^{(2)}=e^{2\delta \rho}(-F(y)\,dt^2+{1\over F(y)}dy^2)}
$${\chi=\chi_{s}+\delta\chi, \qquad f=f_{s}+\delta f}$$
$${(\delta {\chi'}+\lambda\delta\rho) \sim e^{-2\lambda y},
            \quad\hbox{ at most}}.$$

\noindent
Notice that $\delta f$ could be much larger than the static part
$f_{s} \sim e^{-2\lambda y}$.
Expanding the constraint equations $\delta S^{(2)} / \delta
g^{01}=0$ and $\delta S^{(2)} / \delta g^{00}=0$ in $e^{-2\lambda y}$,
and collecting the leading terms,

\eqn\moment{\partial_{t}(\delta \chi'+\lambda\delta\rho)={1\over2}
            \,\delta(f' \dot f),}
\eqn\hamilt{(\delta\chi'+\lambda\delta\rho)'+2\lambda
            (\delta\chi'+\lambda\delta\rho)={1\over 4}
            \,\delta (f' f' +\dot f \dot f).}

\noindent
If $\delta f \sim e^{-\lambda y}$ then $(\delta\chi'+\lambda\delta\rho)
\sim e^{-2\lambda y}$. If $\delta f$ is exponentially smaller than
$e^{-\lambda y}$, the leading part of
$(\delta\chi'+\lambda\delta\rho) \sim e^{-2\lambda y}$
is static and the subleading time-dependent part of it is again
exponentially smaller that $\delta \dot f$.
Finally, the possibility of $\delta f$ much larger than
$\sim e^{-\lambda y}$ is excluded since it,
combined with \moment, will lead to
infinite ADM mass \ADM. After redefining the
unperturbed fields to include the static parts of $\delta\chi,\cdots$,

\eqn\asympt{(\delta\chi'+\lambda\delta\rho) \sim e^{-\lambda y}\delta f
           \qquad\hbox{at most.}}

\noindent
Therefore it is quite reasonable to expect $\delta\chi\,\delta f
\ll \delta f$ as far as the time-dependent perturbations are concerned.
The field equation for $f$ \FFF\ can be expanded in the same fashion
and up to the leading order $\delta f$ is easily shown to be
decoupled from the rest.

\eqn\KGapp{\nabla^2 (\delta f)-2\nabla\chi_{s}\nabla(\delta f)=0}

\noindent
After a field redefinition $\Psi \equiv e^{-\chi_{s}}\delta f$, \KGapp\
becomes,

\eqn\KGAPP{\nabla^2 \Psi- \lambda^2 \Psi=0.}

\noindent
The gravitational perturbation is asymptotically
described by a massive Klein-Gordon equation.

Now consider solutions to this equation with the following characteristic
data, in a flat\foot{\rm The asymptotic geometry is flat within
the same approximation} light-cone coordinate, $g^{(2)}=-dvdu$.

\eqn\leftP{\Psi{\biggr{\vert}}_{v=v_{o}}=0}
$${\psi\equiv\Psi{\biggr{\vert}}_{u=u_{o}}=0
            \quad\hbox{for}\quad v<v_{o},\qquad C(v-v_{o})
            \quad\hbox{for}\quad v_{o}<v<v_{n}}.$$

\noindent
Since the geometry is flat, we can integrate the equation explicitly
using the following recursive solution

\eqn\solP{\Psi(v,u)=-{\lambda^2\over 4}\int^{u}_{u_{o}}\int^{v}_{v_{o}}
          \Psi(v',u')\,dv'du'+ \psi(v).}

\noindent
This uniquely determines $\Psi$ up to $v<v_{n}$ in terms of a
Bessel function.
\eqn\trivial{\partial_{v}\Psi=0\qquad v<v_{o}}
$${\partial_{v}\Psi=C\,J_{0}\bigl ( \lambda\sqrt{(u-u_{o})
            (v-v_{o})}\bigr )\qquad v_{o}<v<v_{n}}$$

\noindent
Notice $[\partial_{v}\Psi]_{v_{o}}=C$.  However the massive nature
of the field becomes important as we move away from $v_{o}$ toward
$v_{n}$. In the future direction ($u>u_{o}$), $\Psi$ is attenuated
and spread in a powerlike manner and as $u \rightarrow \infty$ it
would take an infinite resolution to see the ever so tiny bump
of $\Psi$ owing to the discontinuous derivative.
Because of this tendency to spread the initial signal,
if we had started with a smooth version of the jump $C$ at $u=u_{o}$,
it would have disappeared without trace for $\lambda\,(u-u_{o}) \gg 1$.
The more striking feature is the behaviour in the past ($u<u_{o}$).
Then the square root is pure imaginary and
\eqn\bondi{\Psi \sim e^{\lambda\sqrt{(u_{o}-u)(v-v_{o})}}}

\noindent
up to some powerlike prefactor. On the past
null infinity $u=-\infty$, $\Psi$ is infinite in the interval $(v_{o},
v_{n})$.

In the asymptotically far region, the $f$ field effectively decouples
and the action \grav\ can be considered as the usual dilaton gravity
coupled nonlinearly to  matter fields $f$ and $K$. Then the corresponding
Einstein equation of the metric can be written with energy-momentum
tensor of these matter fields on the right-hand side. $\delta g^{vv}$
equation is given by the following. We did not write down the left-hand
side explicitly.

\eqn\einst{{\delta S^{grav} \over \delta g^{vv}}=e^{-2\chi}
           \partial_{v} f \partial_{v} f+\cdots}
 $$S^{grav} \equiv \int \sqrt{-g^{(2)}} e^{-2\chi}(R^{(2)}
   +4\,(\nabla \chi)^2+4\lambda^2).$$

\noindent
But the leading contribution to the right-hand side
is proportional to the square of \trivial\
which is infinite on past null infinity in the finite interval $(v_{o},
v_{n})$. Physically this means an infinite amount of energy
is sent in from null past infinity  and subsequently the Bondi mass must
be also infinite in the same interval.\foot{\rm Unfortunately
we were not able to derive the mass formula appropriate for the
asymptotic behaviour of $\delta f \sim e^{-\lambda y}$ But if we naively
extrapolate \ADM\ to this case and evaluate on the past null infinity,
the constraint equations \moment\hamilt\ imply infinite Bondi mass.}
(The reason the right hand side of \einst\
is finite for $u>-\infty$ in spite of the initially infinite value can be
understood again by considering the dispersive effect of the
massive dynamical equation. After a sufficient amount of time the
initial flux of infinite density and infinite total energy becomes
a flux of infinite energy yet of finite density spread all the way to
infinite future.)

We can conclude that solutions of the type \trivial\
should be excluded energetically since it involves an infinite amount
of energy. Furthermore the same conclusion holds if we drop the assumption
of  initially static spacetime. Since the perturbation equation is linear
asymptotically, we can subtract a smooth part from $\Psi$ to reduce the
problem to that of an initially static spacetime (i.e., $\Psi=0$ before the
shock).

\vfill\eject
\listrefs
\bye